\documentclass[journal]{IEEEtran}
\usepackage[cmex10]{amsmath}
\usepackage{graphicx}
\usepackage[colorlinks=true, allcolors=blue]{hyperref}
\usepackage[justification=centering]{caption}
\begin{document}
\title{An Analysis of Data Transformation Effects\\on Segment Anything 2}

\author{Clayton~Bromley, Alexander~Moore, Amar~Saini, Doug~Poland, Carmen~Carrano%
\thanks{Clayton Bromley and all authors work in affiliation with\\Lawrence Livermore National Laboratory, Livermore CA\\Email: bromley1@llnl.gov}%
}

\maketitle
\thispagestyle{empty}

\begin{figure*}[h]
    \centering
    \includegraphics[width=0.9\linewidth]{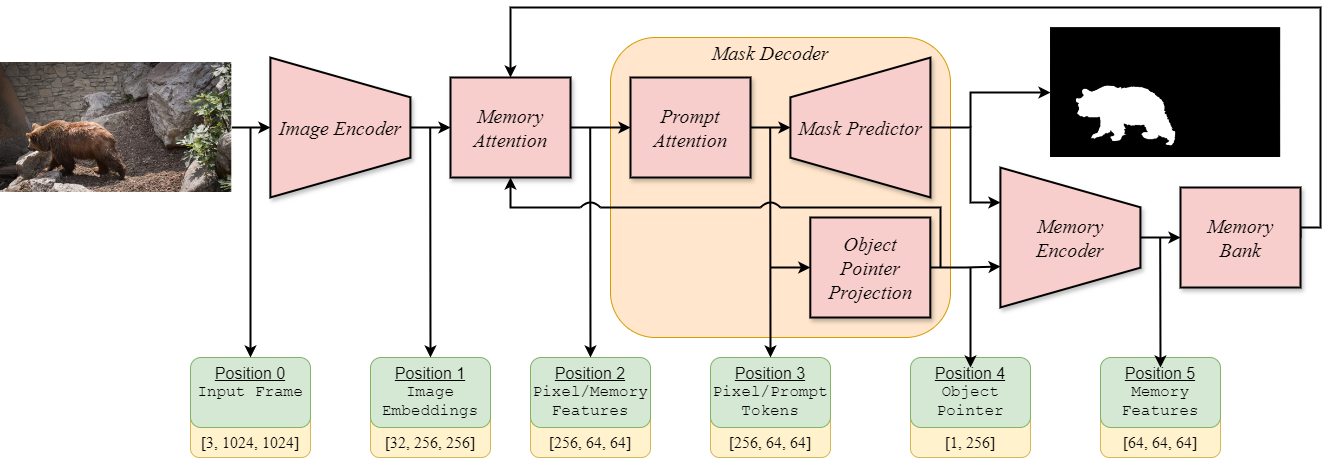}
    \caption{\label{fig:architecture}SAM 2 Architecture and Observation Positions}
\end{figure*}

\begin{abstract}
Video object segmentation (VOS) is a critical task in the development of video perception and understanding. The Segment-Anything Model 2 (SAM 2), released by Meta AI, is the current state-of-the-art architecture for end-to-end VOS. SAM 2 performs very well on both clean video data and augmented data, and completely intelligent video perception requires an understanding of how this architecture is capable of achieving such quality results. To better understand how each step within the SAM 2 architecture permits high-quality video segmentation, a variety of complex video transformations are passed through the architecture, and the impact at each stage of the process is measured. It is observed that each progressive stage enables the filtering of complex transformation noise and the emphasis of the object of interest. Contributions include the creation of complex transformation video datasets, an analysis of how each stage of the SAM 2 architecture interprets these transformations, and visualizations of segmented objects through each stage. By better understanding how each model structure impacts overall video understanding, VOS development can work to improve real-world applicability and performance tracking, localizing, and segmenting objects despite complex cluttered scenes and obscurations.
\end{abstract}


\section{Introduction}

\IEEEPARstart{I}{n} determining the quality of VOS models, benchmarking and quantitative metrics can only tell so much. Although SAM 2 mask predictions are high quality using these metrics, they give little information about how the model perceives objects. By observing how a model reacts to changes in input, it can be determined how perception of VOS models like SAM 2 occurs. Each stage within the SAM 2 architecture alters object perception and, using a series of data transformations, it becomes evident where the model transitions from scene-oriented perception to object-oriented perception.\\

Understanding model perceptions can inform architecture decisions, loss decisions, and potential weaknesses. SAM 2 has flaws, including a weak tracking ability \cite{SAM2Flaw}. By probing SAM 2 with challenging data, potential problems can be identified and areas of improvement can be located. Identifying the layers of SAM 2 is vital to developing future improvements to video object segmentation models.\\

SAM 2's modular design makes it easy to introduce observational positions throughout the model at the image encoding, cross-attention with memory for enriched image features, and cross-attention with sparse prompts for object attention refinement. Tracking embeddings during these observational positions with the introduction of complex datasets designed to test the visual perception of models will enable insight into solving common challenges of designing VOS models for cluttered complex scenes and long-term obscurations.\\

Contributions are as follows.
\begin{enumerate}
    \item The creation of a variety of video datasets with complex transformations, including object obscuration, frame interjections, and object removal.
    \item An analysis of how each stage of the SAM 2 architecture interprets these complex transformations.
    \item A visualization of each observational location, highlighting the features on which the model is focused.
    \item A novel approach to decode SAM 2 object pointers to spatio-temporal representations, demonstrating for the first time how this representation is used by the model.
\end{enumerate}

\section{Related Works}

\subsection{Image Segmentation}\label{sec:image_segmentation}

The SAM 2 architecture is based on the original Segment Anything Model \cite{SAM} (SAM) for image segmentation. When given an input image and a prompt, the model predicts segmentation masks for the associated object, providing precise outlines that can be adjusted or refined based on user input or specific criteria. It operates by leveraging a large dataset of images and their corresponding segmentation masks, allowing it to learn a variety of object shapes and boundaries during an iterative process of annotation and refinement \cite{SAM2}. SAM employs a transformer-based architecture, which enables it to process images in a way that captures both local and global context. SAM architecture uses a heavy-weight image encoder and a light-weight prompt attention mechanism to attend to regions of an image or natural language prompts. SAM returns a set of segmentation masks given the cross-attention of the prompts with the image embeddings, passed through a mask decoder (Figure \ref{fig:architecture}).

\subsection{Video Segmentation}

Until recently, state-of-the-art VOS models, including XMem \cite{XMEM} and Cutie \cite{CUTIE}, relied on the relationship between current and previous frames to predict the relationship between the masks, a process known as frame-to-frame affinity. The semi-supervised VOS structure relies on a single mask prompt and its previous predictions to make the full list of mask predictions. SAM 2 \cite{SAM2} introduced an architecture that is not dependent on affinity, fundamentally changing the way VOS is explored as a problem. Rather than using frame-to-frame affinity for mask prediction, SAM 2 solely utilizes attention between the frame embeddings and the memory. Masks, bounding boxes, and points can all be used as inputs at any point through a video, and prompt encodings are used to decode the mask prediction. SAM 2 applies SAM iteratively and includes the attention of previous frames as a self-prompting mechanism.\\

A simplified architecture diagram for SAM 2 is shown in Figure \ref{fig:architecture}. First, an input image is preprocessed and passed through an image encoder to find the image embeddings. These embeddings undergo two cross-attentional layers, first with the memory of previous mask/frame relations and second with prompt inputs, such as masks, bounding boxes, and point clicks. Mask predictions are decoded as functions of these cross-attentional layers. A 256-dimensional object pointer represents the current state of the object in relation to previous frames. The mask prediction and object pointer are encoded into memory and used for future predictions.

\subsection{Datasets}

Existing datasets explore the segmentation of partially or fully obscured objects. DAVIS \cite{DAVIS} is a dataset of multi-object videos which have been segmented for ground truth masks. It is a highly popular benchmark on which models can compare performance. Although these data contain limited object obscurations, they provide a good baseline for VOS. The DAVIS dataset will be used to generate the data used in this paper.\\

MOSE \cite{MOSE} is a popular video dataset similar to DAVIS that emphasizes object obscuration. In many samples, obscured objects will become partially or fully obscured in a section of the video to test the robustness of the VOS model. State-of-the-art models, particularly Cutie and SAM 2, have emphasized this dataset when developing their architectures and have great performance on these obscurations. The created datasets contain synthetic obscurations that are unreliable when clean data contain natural obscurations. For this reason, DAVIS is used to create a dataset of videos with the necessary annotated long-term interjections.

\section{Experimental Design}

Various complex transformations on the DAVIS dataset are used to collect the data in this paper. The impact of these transformations are observed at five different locations within the architecture as shown in Figure \ref{fig:architecture}.

\subsection{Data Transformations}

While the DAVIS video dataset is used as a baseline for all observed parameters, five complex transformations are applied to create new 500-video datasets. Stochastic resampling of DAVIS videos allows for random sample generation. Object of interest is defined as the object which will be segmented, and obscuring object as a different object that is partially or fully overlapping the object of interest, and the context as the frame background that is not a part of either object.\\

\subsubsection{Four-Frame Interjection}

\begin{figure}[h]
\includegraphics[width=1\columnwidth]{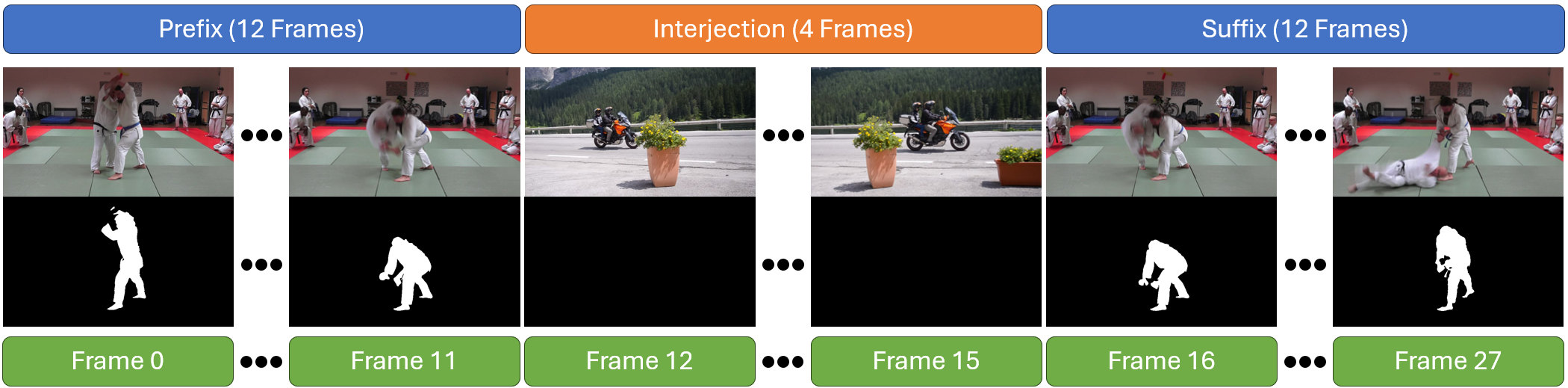}
\caption{\label{fig:interjection_video}Interjection Video Data Structure}
\end{figure}

An object is selected from a video of interest and a select number of frames from a separate, unrelated video are interjected in the middle of the video of interest \cite{Interjection}, as shown in Figure \ref{fig:interjection_video}. Thus, the object of interest is not present during the interjection period. SAM 2 is expected to do the following during the three stages of the interjection video:

\begin{enumerate}
    \item \textit{Prefix}: Take the initial mask and write the object into memory
    \item \textit{Interjection}: Identify the absence of the object and make no false positive predictions
    \item \textit{Suffix}: Reidentify the object and continue segmenting at the same accuracy as during the prefix.
\end{enumerate}

The dataset uses twelve-frame prefixes and suffixes to allow ample time for the model to write the object of interest into memory. Four-frame interjections are used, as shown in Figure \ref{fig:interjection_video}. This dataset tests how resilient SAM 2 is to irrelevant frames. By default, SAM 2 writes to memory every 7 frames \cite{SAM2}. A bad segmentation model may write the interjection frames into memory while making false positive predictions. This would reduce performance in suffix reidentification, as memory is tarnished \cite{Interjection}.

\subsubsection{Object Removal}

\begin{figure}[h]
\includegraphics[width=1\columnwidth]{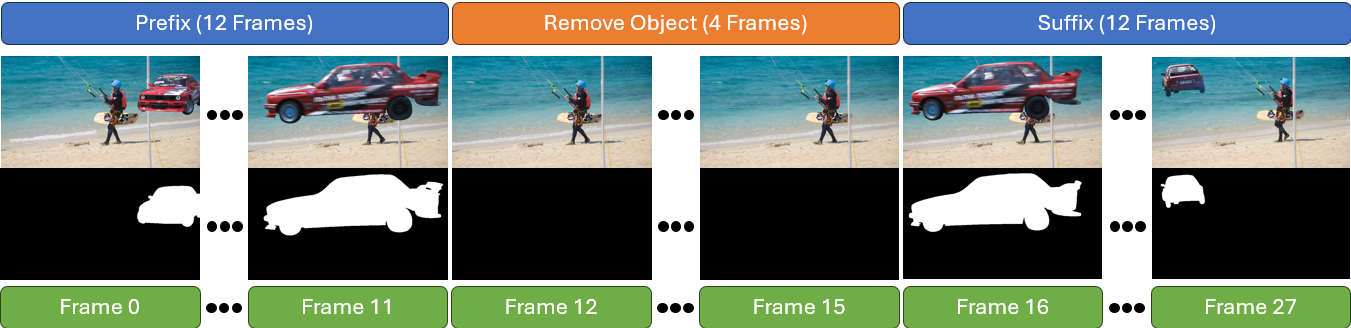}
\caption{\label{fig:object_removal}Object Removal Video Data Structure}
\end{figure}

Like the four-frame interjection dataset, the object removal dataset has a period of time in which the object of interest completely disappears. However, in this dataset, the background context of the video remains. The structure of this dataset is shown in Figure \ref{fig:object_removal}. An object is artificially added to the prefix and suffix of a DAVIS video. This allows this new object to be segmented and than removed without impacting the background context. This dataset will allow for an analysis of how each stage of SAM 2 interprets the object as separate from the surrounding context.

\subsubsection{Context Removal}

\begin{figure}[h]
\centering
\includegraphics[width=1\columnwidth]{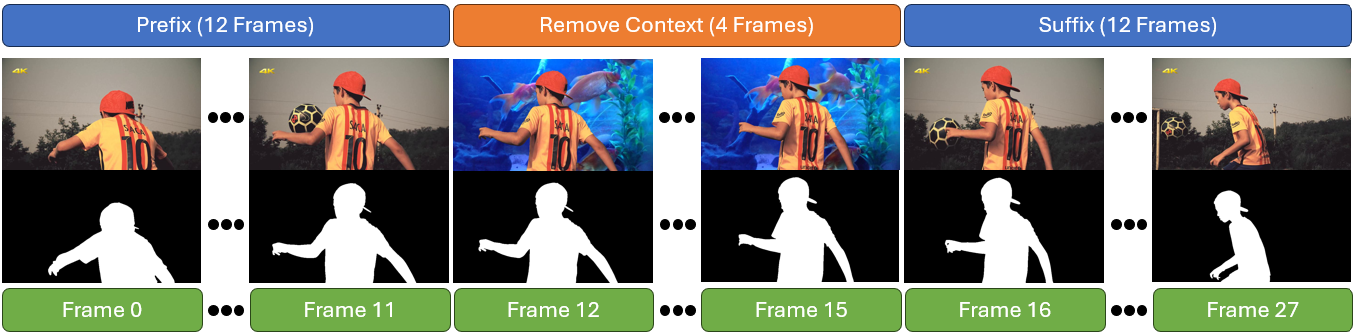}
\caption{\label{fig:context_removal}Context Removal Video Data Structure}
\end{figure}

The context removal dataset is the opposite of the object removal. Rather than the object of interest being removed while the surrounding context remains, the surrounding context is removed. This structure is shown in figure \ref{fig:context_removal}. Unlike the previous datasets, there is no need for object re-identification since the object never disappears. Rather, the model must know how to keep track of the object throughout the entire video. This dataset can test how SAM 2 interprets a large contextual shift that should ideally not affect the segmentation mask predictions. Because frame-to-frame pixel similarity (and thus affinity) is small as the context is removed, this presents a potential failure point for affinity-based algorithms \cite{Interjection}.

\subsubsection{Obscuration}

\begin{figure}[h]
\centering
\includegraphics[width=0.7\columnwidth]{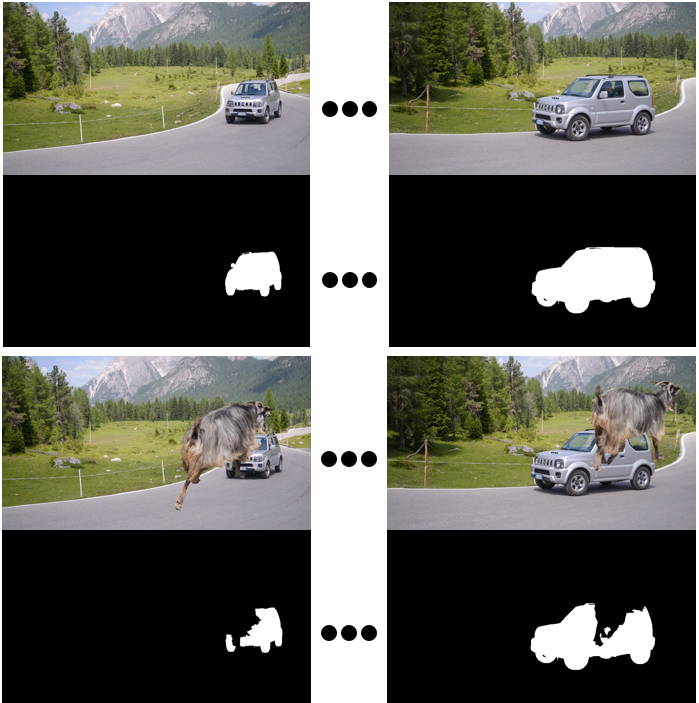}
\caption{\label{fig:obscuration}Obscuration Video Data Structure}
\end{figure}

The obscuration dataset explores the effect of artificially introduced obscuring objects. Figure \ref{fig:obscuration} shows an example of this artificial obscuration as well as the original unobscured DAVIS video. Obscurations of the object of interest present a potential problem of disjointed segmentation areas as well as rapid changes in mask shape from frame-to-frame. This dataset will test how much of an object of interest must be visible for SAM 2 to have an understanding of the object's position and shape.

\subsubsection{Three-Object Overlay}

The 3 object overlay dataset explores the effect of introducing multiple visually similar objects on top of the segmented object. Figure \ref{fig:obscuration} shows an example of this artificial overlay as well as the original unobscured DAVIS video. This dataset will test how well each stage of the SAM 2 architecture can distinguish between several visually similar objects.

\begin{figure}[h]
\centering
\includegraphics[width=0.7\columnwidth]{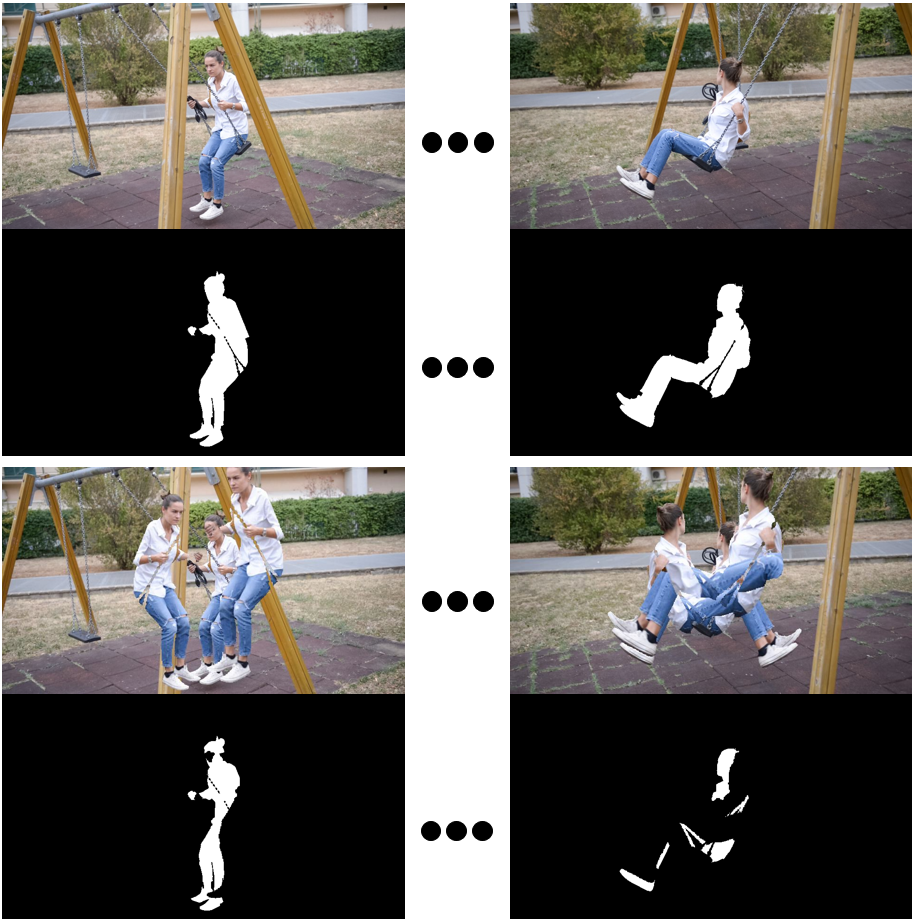}
\caption{\label{fig:nobject}3 Object Video Data Structure}
\end{figure}

\subsection{Observational Positions}

There are five different observational positions within the SAM 2 architecture in which the effects of data transformations are explored. These are shown in Figure \ref{fig:architecture}. In this paper, position 0 refers to the reshaped frame that is initially passed into the model.\\

\subsubsection{Position 1: Image Embeddings}

As with all VOS models, the SAM 2 image encoder creates an embedding representation of the input frame that can be used in later stages. The output of the image encoder is a [32, 256, 256] tensor. As the first step in the SAM 2 architecture, an analysis of this observational position will demonstrate how beneficial initial frame embedding is to overall separation of the object of interest from the surrounding context.\\

\subsubsection{Position 2: Pixel/Memory Features}

Image embeddings undergo cross-attention with the memory of previous frame/mask relations embedded into memory. This stage is used to instill temporal consistency over the mask predictions and ensure that the new predictions follow the same type of object from frame-to-frame. The pixel/memory features output from the memory cross-attentional layer are a [256, 64, 64] tensor. As the primary difference from previous state-of-the-art VOS models, an analysis of the memory cross-attention will show how the affinity-free algorithm interprets past frames in the current mask prediction.\\

\subsubsection{Position 3: Pixel/Prompt Embeddings}

After establishing temporal consistency, the features establish spatial consistency via cross-attention with the input prompts, such as masks, bounding boxes, or point prompts. This process is identical to the original SAM model for image segmentation. The resulting pixel/prompt embeddings are a [256, 64, 64] tensor. This stage will demonstrate how prompt embeddings affect the model's understanding of object localization.\\

\subsubsection{Position 4: Object Pointer}

The output of the SAM 2 mask decoder is a binary mask prediction as well as a [1, 256] object pointer vector which is an embedding representation of the current location of the object. The object pointer is used by future iterations to ensure consistency of object location between each prediction. While the object pointer is embedded into memory, it is also directly used in the first cross-attentional layer to provide insight into previous object locations. An analysis of the object pointer will provide insight into the spatial understanding of the object location with relation to the context.\\

\subsubsection{Position 5: Memory Features}

The mask predictions are object pointers are passed through a memory encoder before they can be stored in memory. These memory features are a useful representation of how SAM 2 remembers each frame for future use. The memory features are a [64, 64, 64] tensor. Understanding what is embedded into memory can provide useful information about what SAM 2 thinks is important for future mask predictions.

\section{Results}

\subsection{Obscuration Datasets}

By visualizing each stage of the SAM 2 process, it becomes evident which elements of a frame are highlighted at each step. To create these visualizations, a channel-wise operation is performed, leaving only a two-dimensional representation of the image spaces. The channel-wise mean and variance are explored. Then each visualization is scaled to the same dimensions and pasted together.\\

\subsubsection{Obscuration}

\begin{figure}[h]
\includegraphics[width=1\columnwidth]{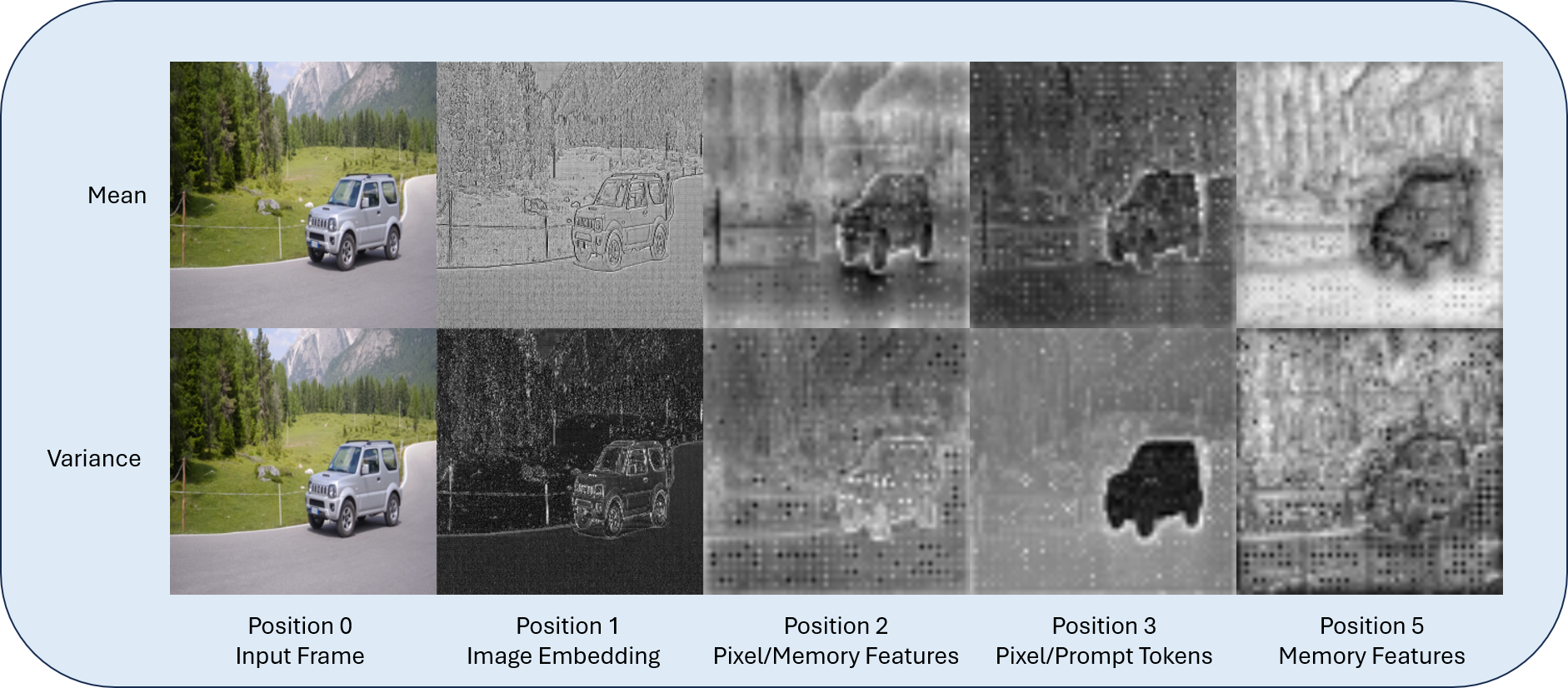}
\caption{\label{fig:ct_vis}Clean DAVIS Video Position Visualizations}
\end{figure}

\begin{figure}[h]
\includegraphics[width=1\columnwidth]{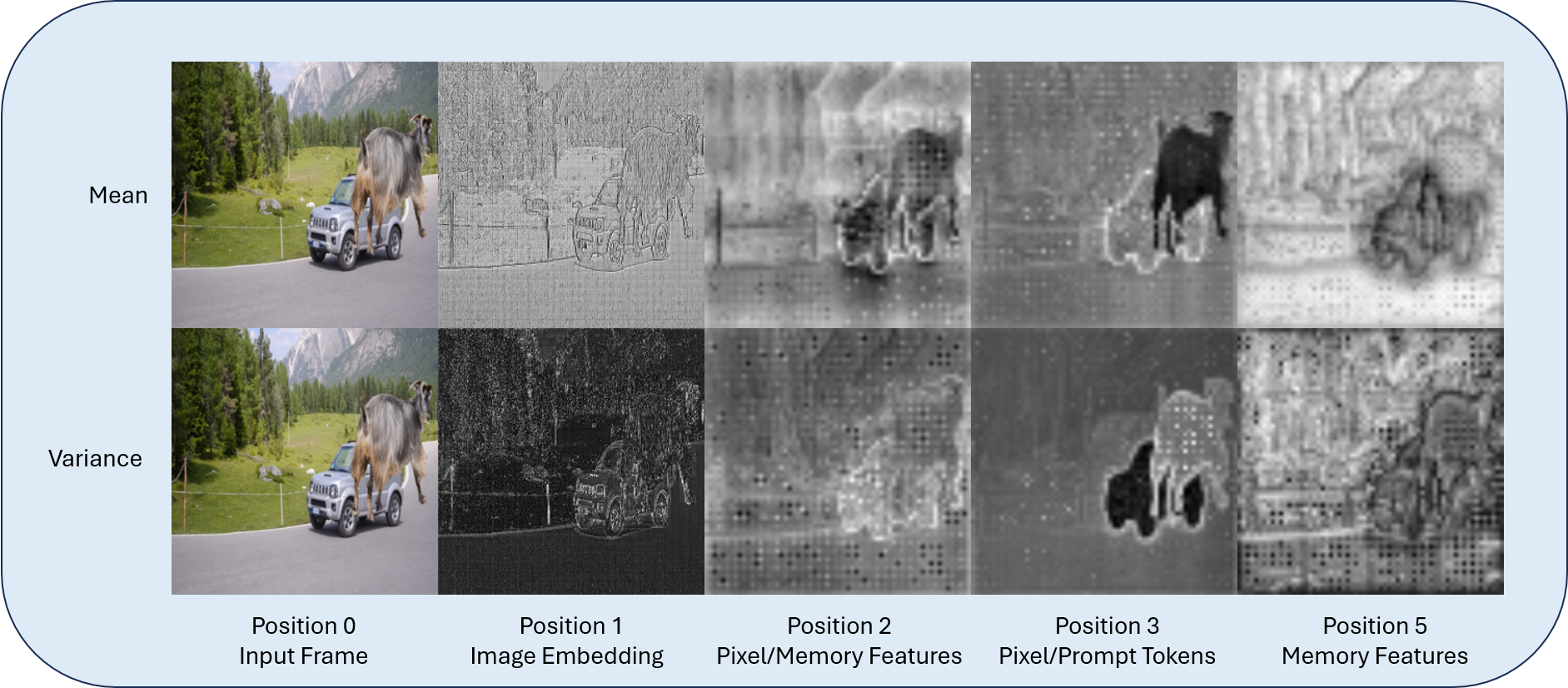}
\caption{\label{fig:ob_vis}Obscuration Video Position Visualizations}
\end{figure}

A visualization of positions 0 (raw image), 1 (image embeddings), 2 (after memory attention), 3 (after prompt attention), and 5 (memory features) from a single frame are shown for a clean DAVIS video in Figure \ref{fig:ct_vis} and for and obscured version of the same video in Figure \ref{fig:ob_vis}. The top row in each figure uses a channel-wise mean to create two-dimensional visualizations while the bottom row shows a channel-wise variance. The car is the object of interest in both videos, while the goat is the obscuring object.\\

The image embeddings in position 1 clearly represent some kind of edge detection. This stage is before attention with the mask history. Thus, as shown in both figures, there is no indication of the object location or shape, rather just a representation of the entire frame. In both cases, the edges have a small mean and a large variance in the channel dimension compared to the rest of the embedding positions.\\

The memory attention embeddings in position 2 take into account the previous frame memory embeddings and the object pointers from previous frames. The car object in the clean video has a small mean with large variance compared to its surroundings, showing that the model is able to identify the object at this stage. In the obscuration video, however, the obscuring object also has a small mean and large variance, demonstrating that the model can not yet distinguish which object is the correct object, just the object location.\\

The prompt attention embeddings at position 3 are the first point at which the model shows an understanding of which is the desired object. The car in both videos has an extremely low variance, and the border pixels have a large mean, creating a glow around the object. The obscuring object has a very low mean and a high variance. It becomes increasingly apparent which pixels belong to the object.\\

Finally, the memory features at position 5 focus less on the object and more on the entire image. The details of the entire frame are visible, and the object of interest is highlighted in a dark glow of low mean and low variance. It is useful for the model to train an understanding of the image background as well as the location of the object, and this is demonstrated in the memory feature visualization.

\begin{figure}[h]
\includegraphics[width=1\columnwidth]{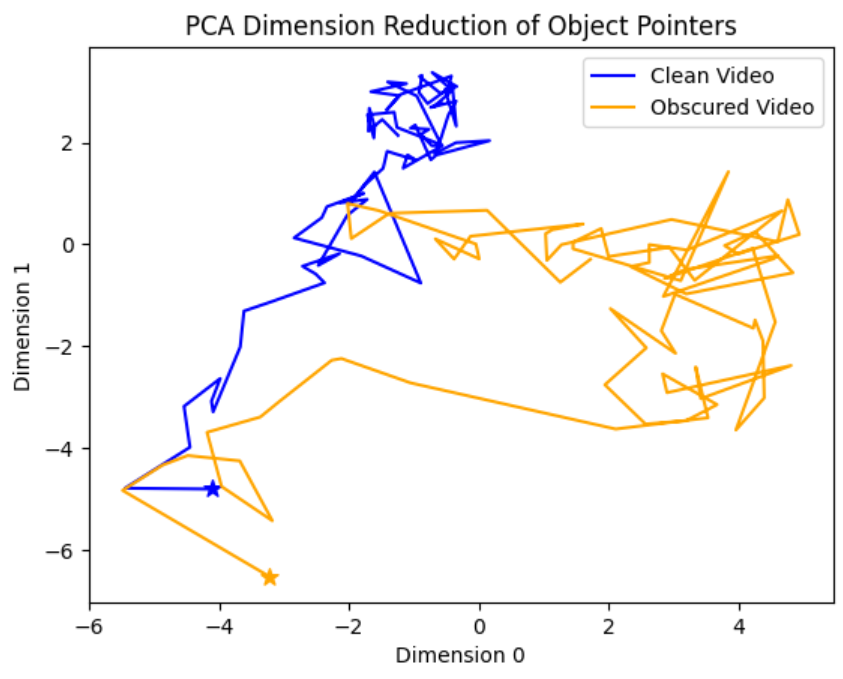}
\caption{\label{fig:ob_pca}Two-Dimenstional PCA Reduction of Obscuration Object Pointers}
\end{figure}

Because the object pointer at position 4 is a vector [1, 256] without explicit spatial and channel dimensions, it is difficult to visualize it the same way as the other positions. As such, it is instead visualized in Figure \ref{fig:ob_pca} using a two-dimensional reduction PCA analysis. The object pointers from all frames of both videos are shown with a star signifying the first frame.\\

In the obscured video, the object of interest (the car) is initially obscured. For most of the video, the obscuring object (the goat) covers part of the car, and the end of the video shows the car once again unobscured. The PCA visualization shows a similar pattern, with the object pointers from the same frame appearing closer together when the object is unobscured and further apart when the object is obscured.\\

\begin{figure}[h]
\includegraphics[width=1\columnwidth]{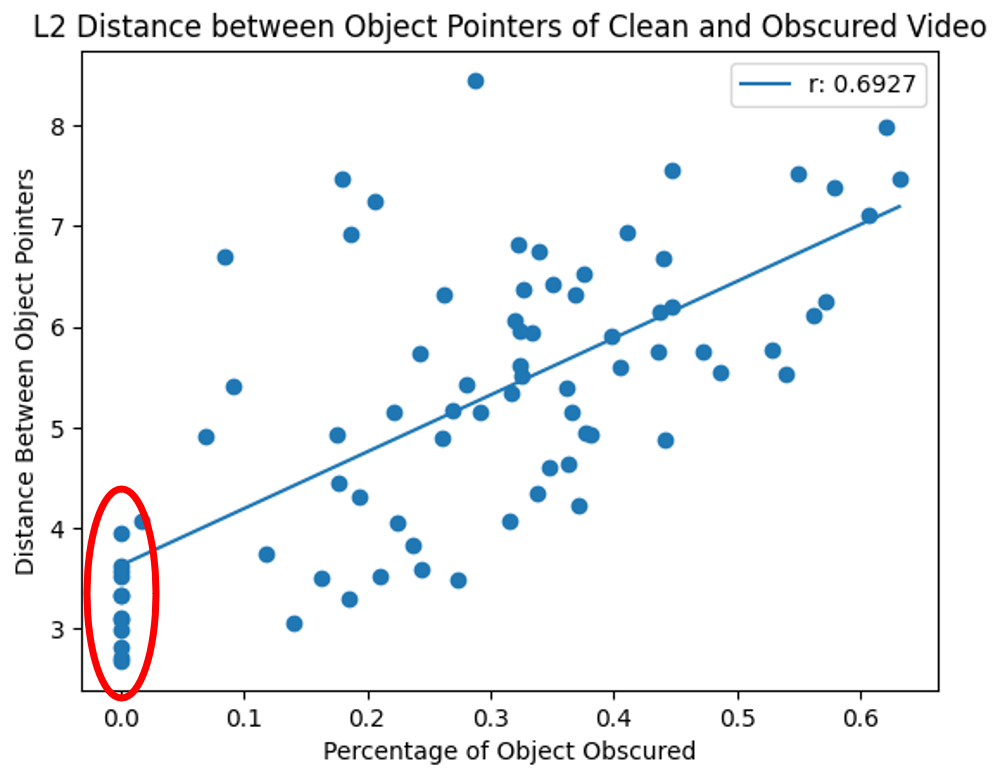}
\caption{\label{fig:ob_dist}Correlation Between Object Obscuration and Distance Between Object Pointers}
\end{figure}

To test this correlation, Figure \ref{fig:ob_dist} shows the L2-distance between the object pointers in the two videos as a function of the percentage of the object that is obscured. These two variables clearly show a positive correlation. The points circled in red show an interesting pattern. All of these frames are 0\% obscured, yet the distance between the object pointers is not constant. This indicates that the object pointer must not just be an embedding object location and shape. Having an obscuring object in the frame changes the object pointer even if the object of interest is still fully visible.

\subsubsection{Three-Object Overlay}

The same visualizations are shown for a three-object dataset sample and the corresponding clean DAVIS sample in Figures \ref{fig:nob_vis} and \ref{fig:sw_vis} below. As with the obscuration sample, the position 1 image embeddings are a simple edge detector where the edges are represented with low mean and high variance in the channel dimension. The position 2 memory attention features show an understanding of the object shape, but not position. All of the overlayed objects are highlighted in the three-object visualization despite representing different objects. The position 3 prompt-attention embeddings allow for a object localization. Only the object of interest, not the overlayed objects, is highlighted with a low variance, indicating that the prompt attention set enforces spatial consistency. Finally, the position 5 memory features are very similar in both the original and transformed visualizations. A representation of the entire frame with a highlight over the object is embedded into memory in both samples.\\

\begin{figure}[h]
\includegraphics[width=1\columnwidth]{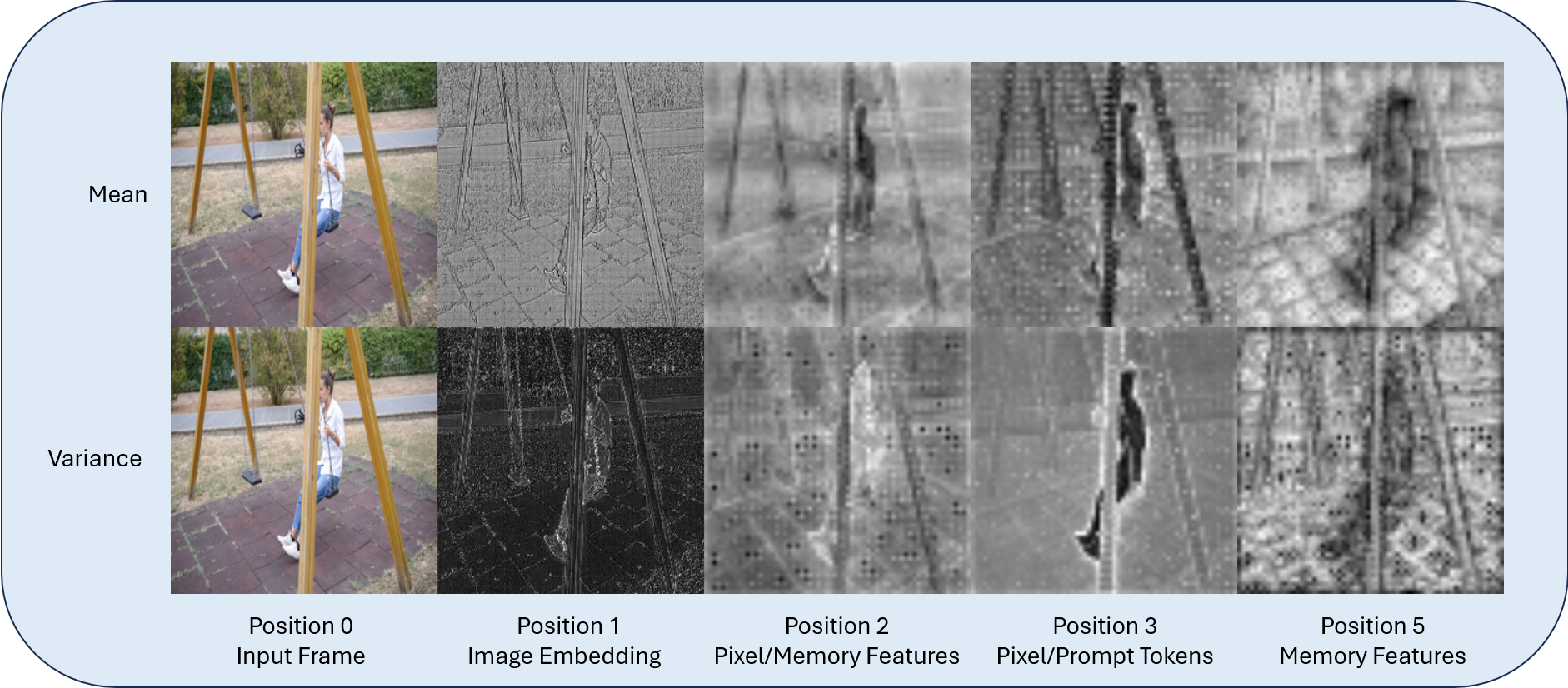}
\caption{\label{fig:sw_vis}Clean DAVIS Video Position Visualizations}
\end{figure}

\begin{figure}[h]
\includegraphics[width=1\columnwidth]{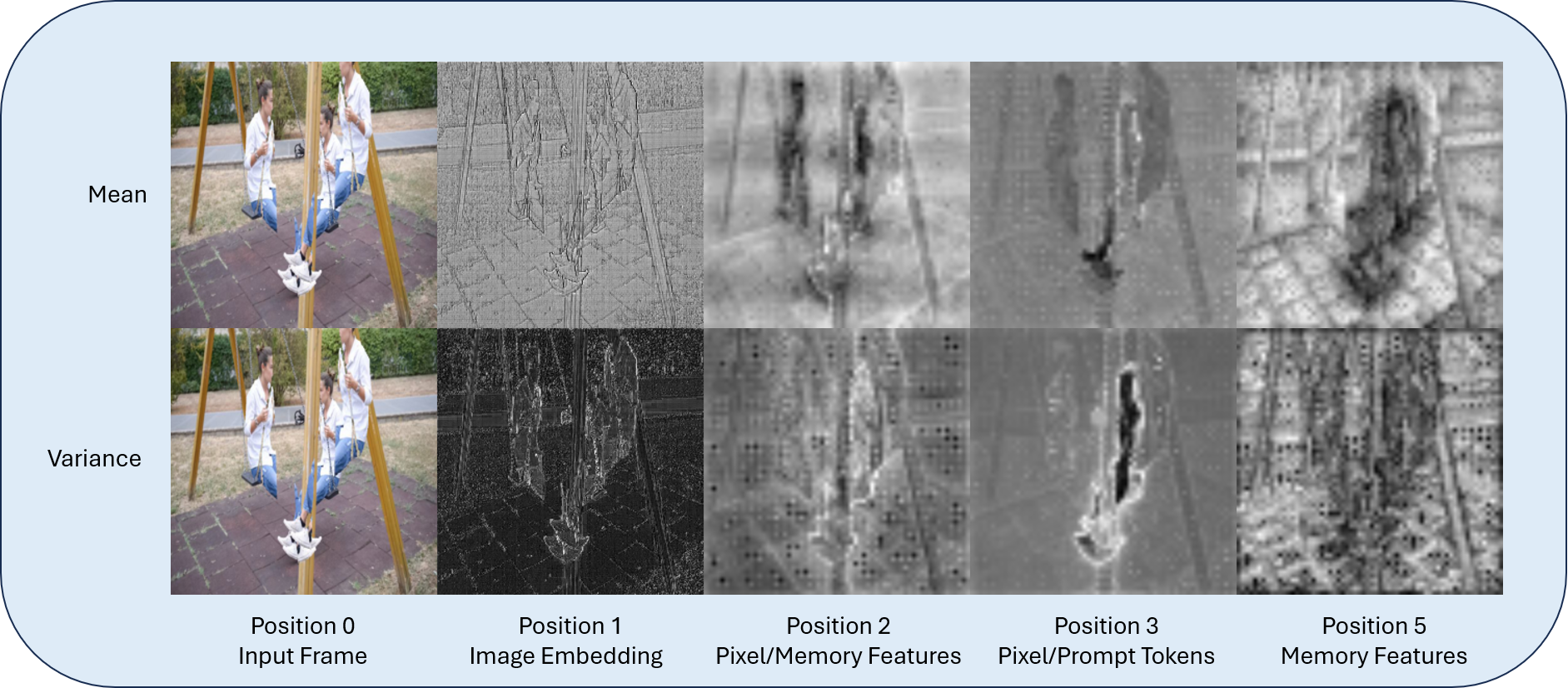}
\caption{\label{fig:nob_vis}3-Object Video Position Visualizations}
\end{figure}

\begin{figure}[h]
\includegraphics[width=1\columnwidth]{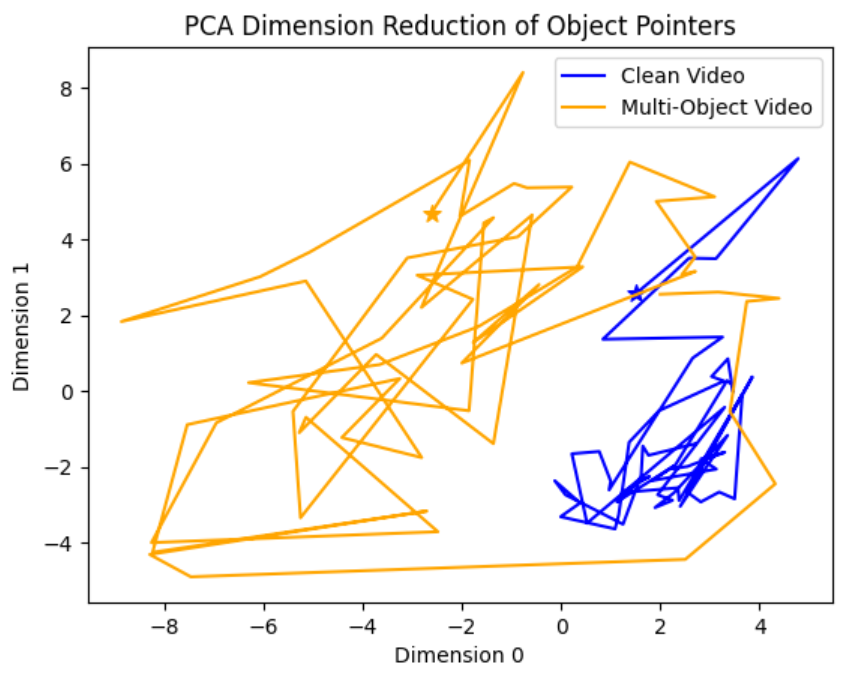}
\caption{\label{fig:nob_pca}Two-Dimensional PCA Reduction of Multi-Object Object Pointers}
\end{figure}

Figure \ref{fig:nob_pca} shows a visualization of the two-dimensional PCA reduction for both samples. The original, clean video dimension values present a smaller cluster than the transformed video. Introducing object overlays causes an increase in the object pointer variance.

\subsection{Object Pointer Analysis}

\begin{figure}[h]
\includegraphics[width=1\columnwidth]{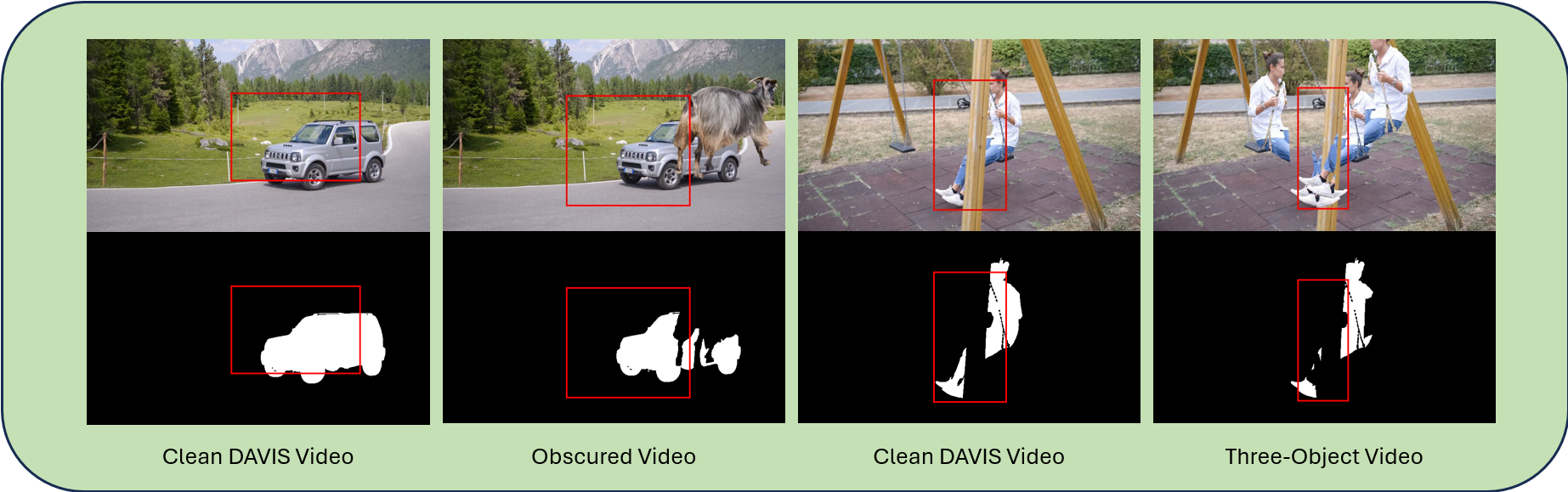}
\caption{\label{fig:obj_ptr}Predicted Object Location using Object Pointer}
\end{figure}

As described in the SAM 2 paper \cite{SAM2}, the object pointer is a [1,256] vector used by the model to recall the position of the object in previous frames. It is created as an embedding of the object location and fed back during cross attention with memory. This raises the question of what exactly the object pointer represents and how it is perceived by the model. To determine if there is a spatial relationship between the object pointers and the object in the original frame, basic models are explored to decode the object position. A basic multilevel perceptron (MLP) model is trained on the entire DAVIS dataset with the [1,256] object pointer input and the ground truth [1,4] object bounding box output, representing [xmin, ymin, xmax, ymax]. This trained model is used to predict object location in the current frame using only the object pointer. Figure \ref{fig:obj_ptr} shows the projection of the object pointer resulting in the same frames shown in Figures \ref{fig:ct_vis}, \ref{fig:ob_vis}, \ref{fig:sw_vis}, and \ref{fig:nob_vis}, respectively.\\

\begin{figure}[h]
\includegraphics[width=1\columnwidth]{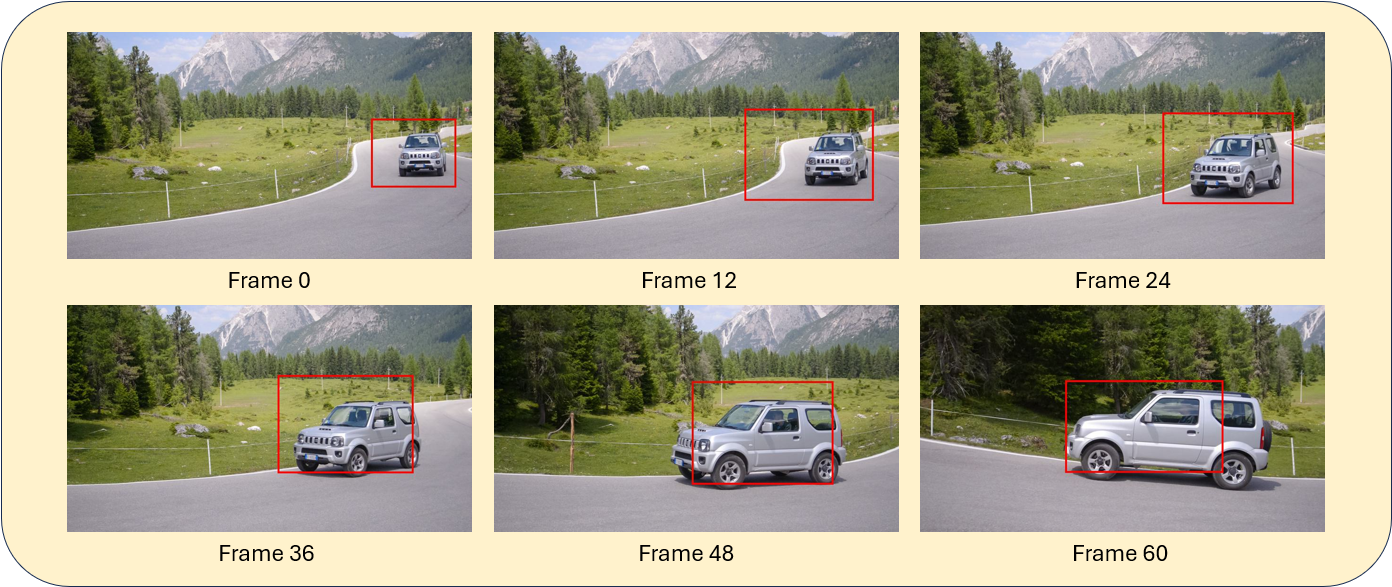}
\caption{\label{fig:ct_op}Object Pointer Projection Visualizations}
\end{figure}

While the resulting bounding box visualizations are not perfect, there is certainly a correlation to the position and size of the object in the frame. This representation allows a proper visualization of where SAM 2 is looking. Figure \ref{fig:ct_op} shows the projection of the object pointer onto a bounding box for several frames from the same video, thus showing that the object pointer is highlighting the car's location over time.\\

These object pointers may have connection to the object velocity within the frame, as the center of the resulting projection tends to be on the leading side of the object motion. This motion is essential to perception and allows the model to focus on the attentional object \cite{MBS}.\\

\subsection{Interjection Datasets}

Interjection datasets refer to the datasets for which there is a four-frame interjection period (four-frame interjection, object removal, and context removal) as well as the clean DAVIS dataset as a baseline. It is important to note that the clean DAVIS and context-removal datasets maintain object presence throughout the entire video while the four-frame interjection and the object removal datasets do not have the object present for frames 12-15. Ideally, the explored features, defined in appendix \ref{methods}, should remain relatively constant for the frames that show the object. The features should increase significantly during the interjection period (frames 12-15) for the four-frame interjection and object removal datasets.\\

\subsubsection{Position 1 - Image Embeddings}

\begin{figure}[h]
\includegraphics[width=1\columnwidth]{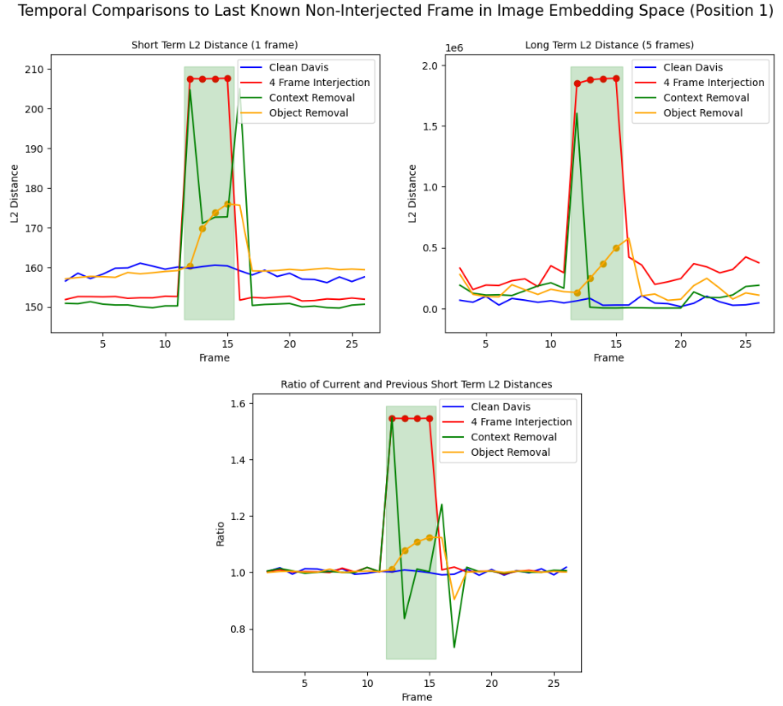}
\caption{\label{fig:pos1_results}Position 1 Features over Time}
\end{figure}

Figure \ref{fig:pos1_results} shows the average over the entire dataset of the observed features through time in the embedding space of the image position 1. For each of the three features, all of the samples of the four interjection datasets are averaged for each frame, such that each line shows the average trend of that feature over each frame. The lines with marked points represent the frame in which the object is not present. This is during the interjection period for the four-frame interjection and object removal datasets. The green shaded region is the interjection region.\\

In all three feature plots, the baseline clean DAVIS dataset is relatively constant, as expected. The metric applied to the four-frame interjection dataset (red line) increases significantly during the interjection period (frames 12-15), as each interjection frame is compared to the prefix frames (frames 0-11). On average, each interjection frame is further from the last known frame containing the object than two frames from the same video are from each other.\\

The metrics applied to the context removal dataset (green line) follows an interesting pattern in the image embedding space. Ideally, these plots will be constant, like the clean DAVIS plots. Because the context is removed during the interjection period, the frames are very distant from earlier frames. The large jump in feature values in frame 12 reflect this frame dissimilarity. However, the object is still present. As such, the subsequent frames are being compared to the removed-context frames, resulting in a lower distance for the remainder of the interjection period. The short-term distance and ratio plots show an additional jump in frame 16 when the context returns.\\

The metrics applied to the object removal dataset (orange line) experiences the inverse effect. These plots should match the four-frame interjection plots with a large jump during the interjection period. Because the background context of the frames in the interjection period remain, the frame is generally similar during this period. This is shown by the lack of a large increase during the interjection period. Instead. the feature values increase slightly, representing the small change in frame caused by the removed object.\\

\subsubsection{Position 4 - Object Pointer}

\begin{figure}[h]
\includegraphics[width=1\columnwidth]{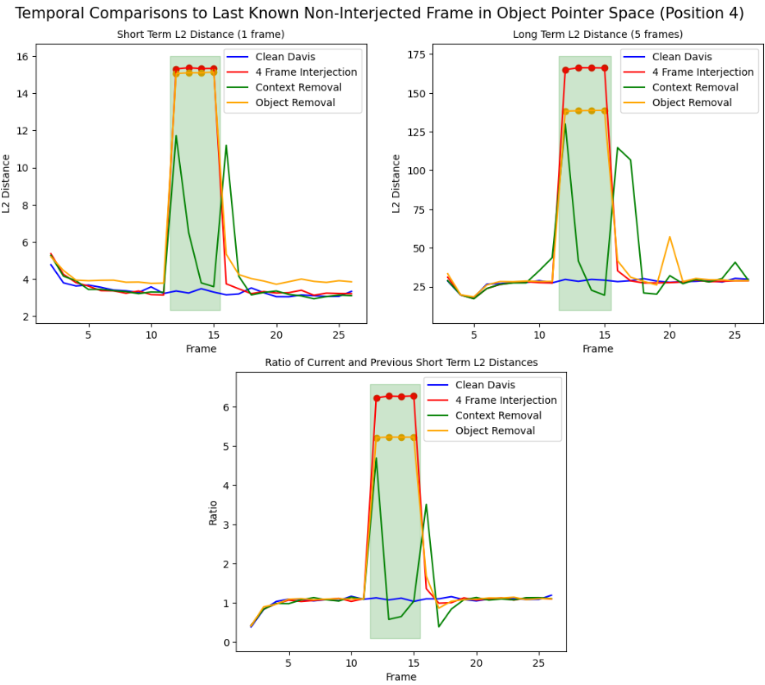}
\caption{\label{fig:pos4_results}Position 4 Features over Time}
\end{figure}

While the previous positions are representations of the entire image, the object pointer is a direct representation of the object. As such, the results of the object features are the cleanest. As shown in Figure \ref{fig:pos4_results}, the object removal dataset is nearly identical to the four-frame interjection dataset plot, although the magnitude may be different. This means that the object pointer is fully connected to the object of interest, rather than the irrelevant background details. While the context removal dataset plot still contains two peaks, the maximum magnitude of these peaks are lower than the object removal dataset during the interjection period.

The object pointer space is the ideal representation of the object separated from its surroundings. Analysis of the object pointer can create a nearly complete picture of the segmented object. Further analysis of the observed features and additional plots for all positions are shown in Appendix \ref{results}

\section{Conclusion}

SAM 2, the current state-of-the-art video object segmentation model, employs a unique architecture that solely relies on cross-attention with memory. In order to improve real-world applicability of tracking, localization, and segmentation models, particularly on complex scenes, it is necessary to understand how each stage of the SAM 2 architecture interprets video. In this paper, Five new datasets with complex transforms were proposed, each testing the robustness of SAM 2 on different challenges. Five different observational positions at each major stage within the model were analyzed, and it was determined at which stage the model is able to discern the object of interest from obscuring objects and the surrounding context. Primarily, the object pointer provides unique insight into where the model is spatially focused when writing to memory and when subsequent frames perform memory cross-attention. By visualizing how each architectural structure impacts the overall video understanding, VOS development can improve real-world applicability despite complex cluttered scenes and video transformations.


%

\appendices
\section{Analytical Methods}
\label{methods}

To analyze the effects of certain complex data transformations, particularly transformations that change throughout the video, various features must be observed to compare frames to the equivalent position in previous frames. The distance between the embeddings at a particular observation position for the current and previous frames is used as this metric. The L2 distance is computed using Equation 1. $f_i$ represents the embeddings of the current frame in stream, $f_m$ represents the element-wise mean of the equivalent embedding matrices for the frames in memory, and $\sigma_m$ represents the element-wise variance of the embeddings of the previous frames. The element-wise difference between the current and previous embedding matrices is found and normalized by the element-wise variance, and the magnitude of these differences is returned.

\begin{equation}
   \text{Regularized L2 Distance} = \left\lvert\left\lvert\frac{f_i - f_{m}}{\sigma_m}\right\rvert\right\rvert
\end{equation}

The reference frame embeddings come from a stored list of embeddings at that observational position that are known to contain the object of interest (i.e. not an interjection frame or an object-removal frame). For a context window of size w, the previous w frame embeddings are used to find the element-wise mean ($f_m$) and element-wise variance ($\sigma_m$) such that $f_i$, $f_m$, and $\sigma_m$ all maintain the same dimensionality. A context window is necessary to avoid overfitting frames with small variances over long video lengths.\\

In real-world applications, it is useful to distinguish frames containing an object from frames in which the object is not present \cite{Interjection}. The computed metric can be used to create a classifier to separate visible frames from artificially interjected frames without the object. In this paper, three features are explored that result from these computed regularized distances that may be used as a classifier. The nature of SAM 2 can be understood by quantifying the differences between clean and transformed videos in the equivalent observational positions.

\subsection{Zeroth-Order L2 Comparison}

A zeroth-order L2 comparison considers the raw computed L2 distance value at a particular window length. In this paper, short-term L2 distance refers to the zeroth-other comparison with a window of length one, such that each frame is directly compared to the previous frame known to contain the object. The long-term L2 distance has a window of length five, such that each frame is compared to the previous five frames known to contain the object.\\

This is the simplest classifier threshold possible. If the distance from a frame to the previous frames at a particular stage in the architecture is above a certain threshold, it is a valid assumption that the object is not visible. If the distance is below the zeroth order threshold, then the object may still be visible.

\subsection{First-Order L2 Comparison}

In certain cases, the most dissimilar frames from the same video may be less similar than two frames from different videos. As such, zeroth-order classification is imperfect. First-order classification looks at the ratio of the current zeroth-order distance to the previous distance. This allows the model to determine if an interjected frame is less similar to the original video than the original video is to itself, thus improving the classification accuracy. In this paper, the short-term L2 ratio as a feature of each stage is explored.

\section{Extended Results}
\label{results}

\subsection{Position 1 - Image Embeddings}

\begin{figure}[h]
\includegraphics[width=1\columnwidth]{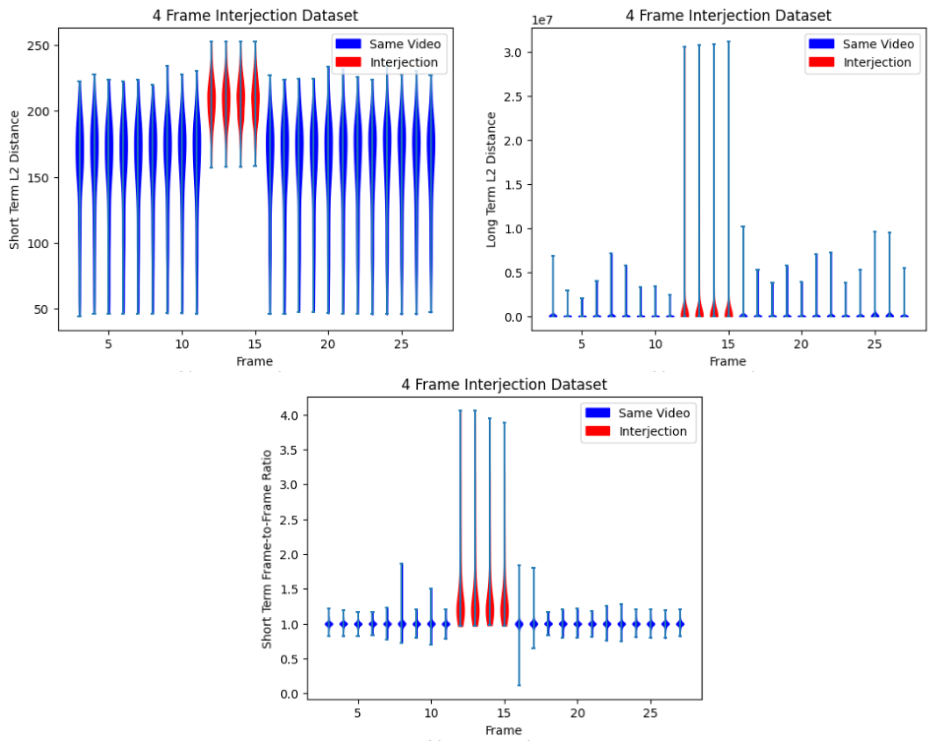}
\caption{\label{fig:pos1_int}Four-Frame Interjection Position 1 Separability}
\end{figure}

\begin{figure}[h]
\includegraphics[width=1\columnwidth]{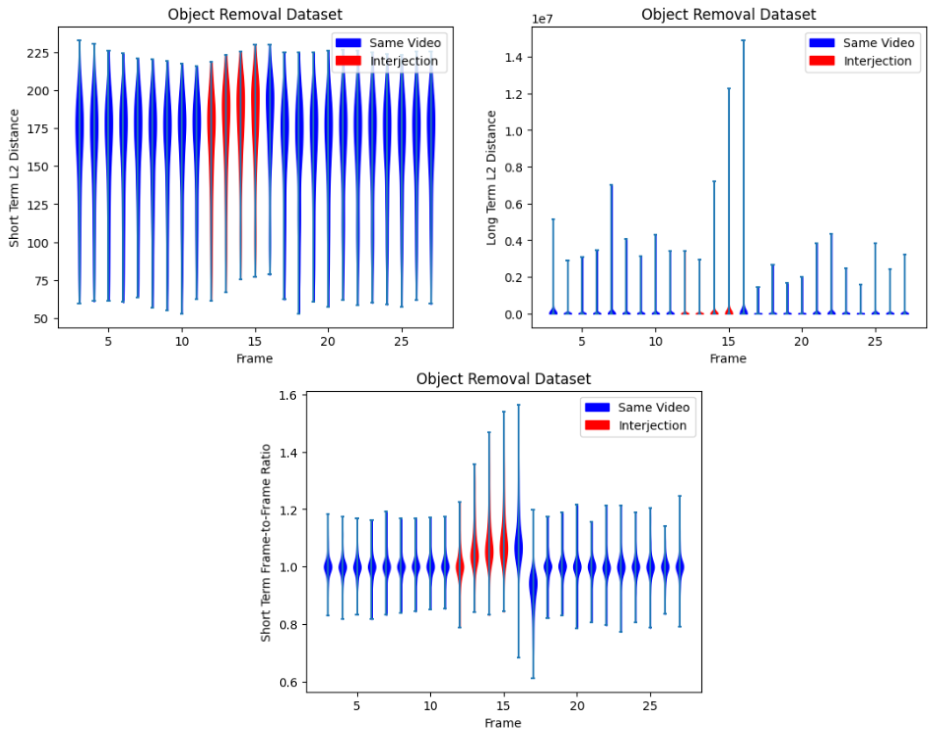}
\caption{\label{fig:pos1_rmo}Object Removal Position 1 Separability}
\end{figure}

While Figure \ref{fig:pos1_results} shows the average over all samples for position 1, Figures \ref{fig:pos1_int} and \ref{fig:pos1_rmo} show the distribution of each feature value for each frame for the four-frame interjection and object removal datasets respectively. The interjection frames that do not contain the object are colored red. In order to design a perfect classifier of the presence of the object in a frame, the red and blue frames must be completely separable. At this stage, the four-frame interjection plot regions are slightly separable and are increasingly separable when using a linear combination of the three features. However, they are not fully separable at this stage. The object removal dataset plot region are completely inseparable at this stage.\\

These plots show that the SAM 2 image encoder does not understand the presence of objects. In this position, the feature values increase only when there is a large change in the raw image, independent of a change in the presence of the object.

\subsection{Position 2 - Pixel/Memory Features}

\begin{figure}[h]
\includegraphics[width=1\columnwidth]{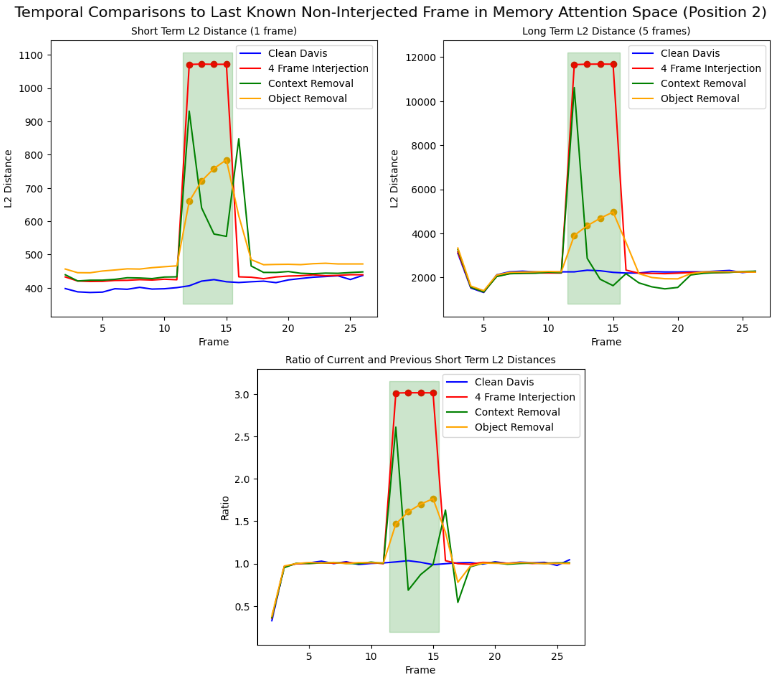}
\caption{\label{fig:pos2_results}Position 2 Features over Time}
\end{figure}

Figure \ref{fig:pos2_results} shows the average feature value for each dataset in each frame after memory attention. The peaks of the context removal dataset line, while still visible, are shorter than in the previous stage. This implies that the model is more often able to recognize that the object is still present during the interjection period. The object removal dataset line is also higher than in the previous stage, implying that the model can recognize that the object is not visible.

\begin{figure}[h]
\includegraphics[width=1\columnwidth]{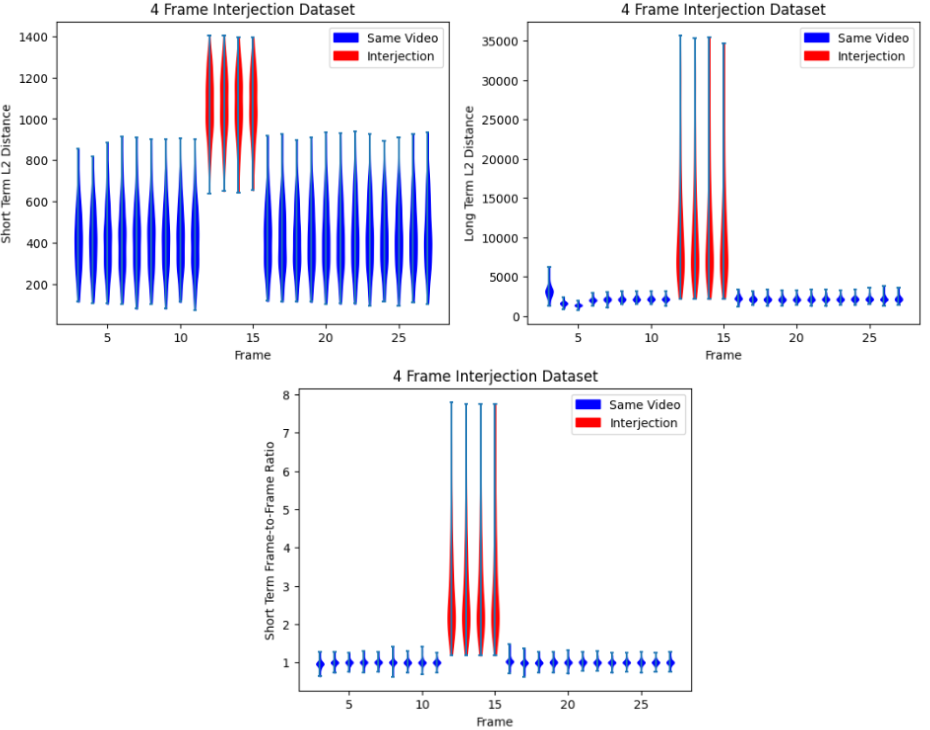}
\caption{\label{fig:pos2_int}Four-Frame Interjection Position 2 Separability}
\end{figure}

\begin{figure}[h]
\includegraphics[width=1\columnwidth]{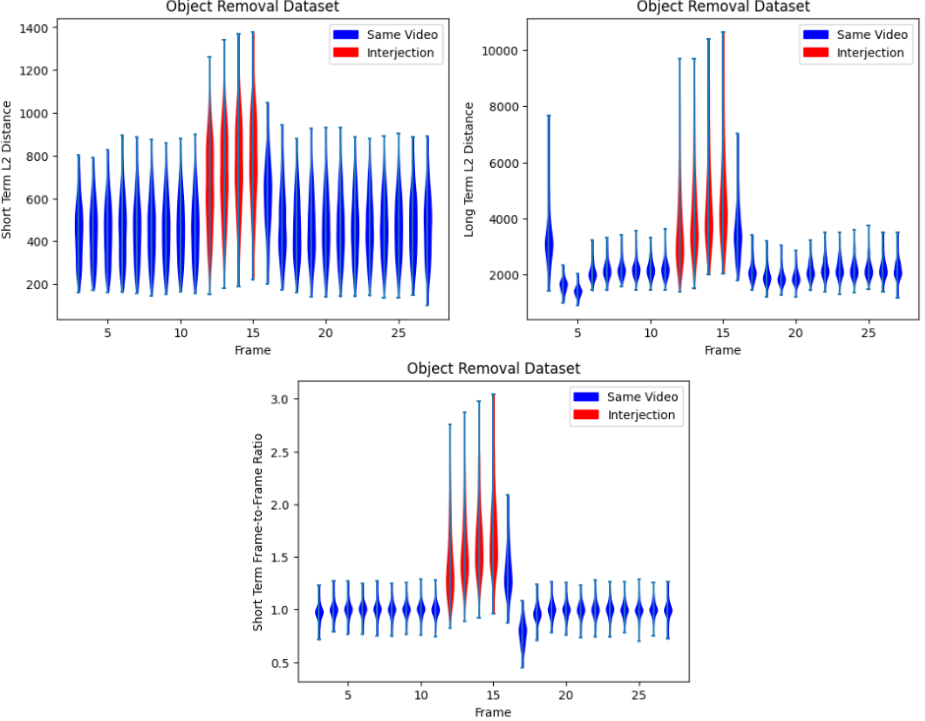}
\caption{\label{fig:pos2_rmo}Object Removal Position 2 Separability}
\end{figure}

Figures \ref{fig:pos2_int} and \ref{fig:pos2_rmo} show the separability of the interjection region for the four-frame interjection and object removal datasets. Although the features are still not fully separable individually, linearly combining these features results in a far more separable solution. As in position 1, the four-frame interjection dataset is much more separable than the object removal dataset, showing that the model is still influenced by the background of each frame rather than just the object.

The memory attention stage of the SAM 2 architecture establishes a temporal consistency in object appearance, and this is reflected in these visualizations. When the object disappears in interjection frames, the model can identify that the objects still present do not match the appearance of the object in previous frames. The integration of previous frame object pointers helps enforce this consistency.

\subsection{Position 3 - Pixel/Prompt Embeddings}

\begin{figure}[h]
\includegraphics[width=1\columnwidth]{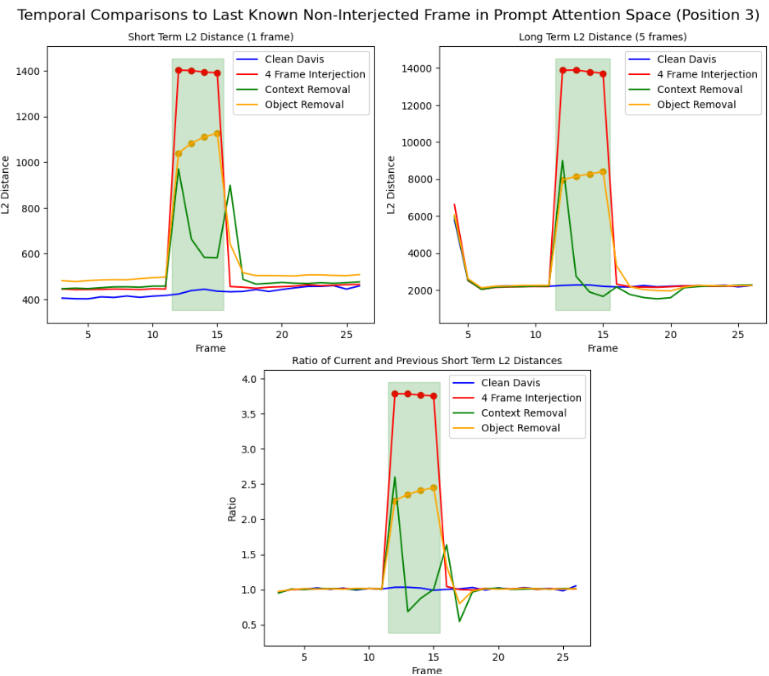}
\caption{\label{fig:pos3_results}Position 3 Features over Time}
\end{figure}

Figure \ref{fig:pos3_results} shows the results from after the prompt attention stage. As with position 2, the model shows an increased understanding of object location. The peaks from the context removal dataset line are shorter than the increase in the object removal dataset plot.\\

\begin{figure}[h]
\includegraphics[width=1\columnwidth]{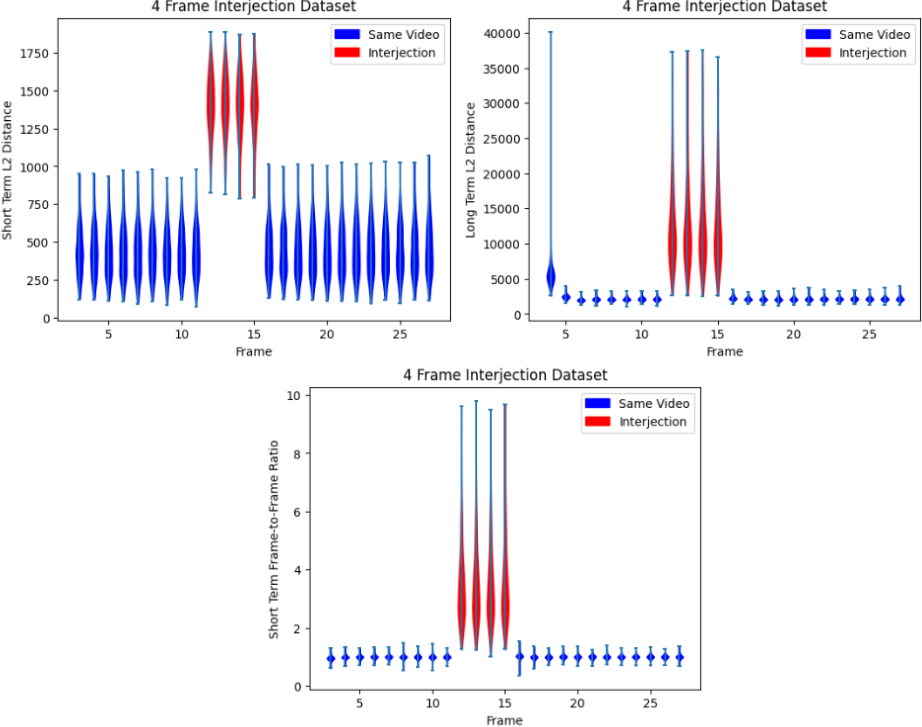}
\caption{\label{fig:pos3_int}Four-Frame Interjection Position 3 Separability}
\end{figure}

\begin{figure}[h]
\includegraphics[width=1\columnwidth]{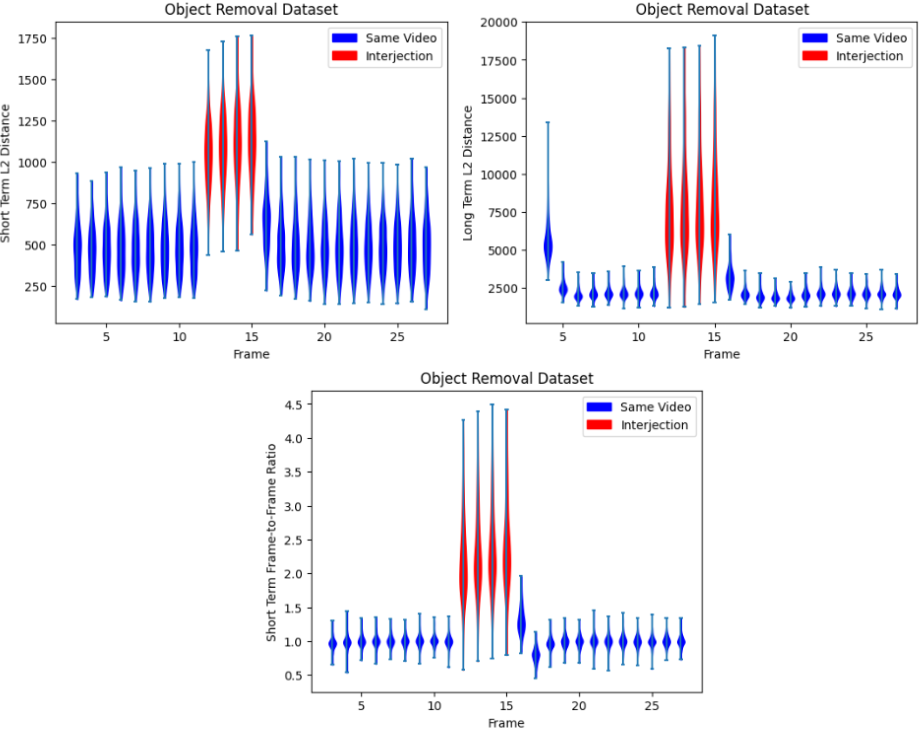}
\caption{\label{fig:pos3_rmo}Object Removal Position 3 Separability}
\end{figure}

In Figures \ref{fig:pos3_int} and \ref{fig:pos3_rmo}, the interjection region is now significantly more separable, and by linearly combining the features, both the four-frame interjection and object removal datasets should be completely separable.\\

The prompt attention layer is used to create spatial consistency between the frames, aligning the position of a predicted object with the position from previous frames. In instances where similar-appearing objects are present in interjection frames, cross-attention in the prompt space will ensure that only the same object will be segmented.

\subsection{Position 4 - Object Pointer}

\begin{figure}[h]
\includegraphics[width=1\columnwidth]{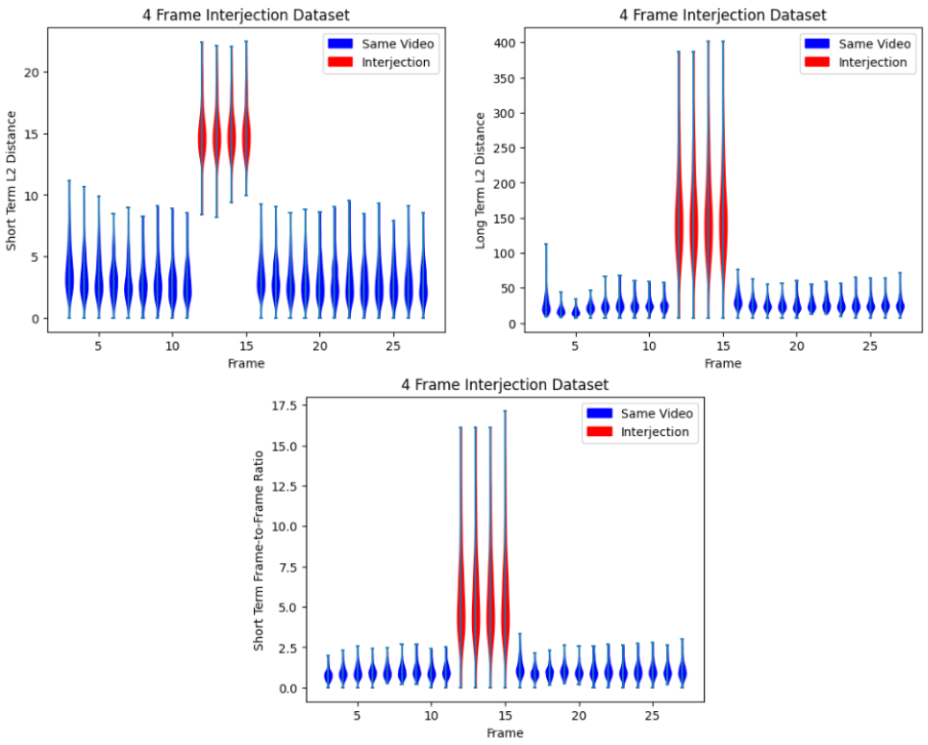}
\caption{\label{fig:pos4_int}Four-Frame Interjection Position 4 Separability}
\end{figure}

\begin{figure}[h]
\includegraphics[width=1\columnwidth]{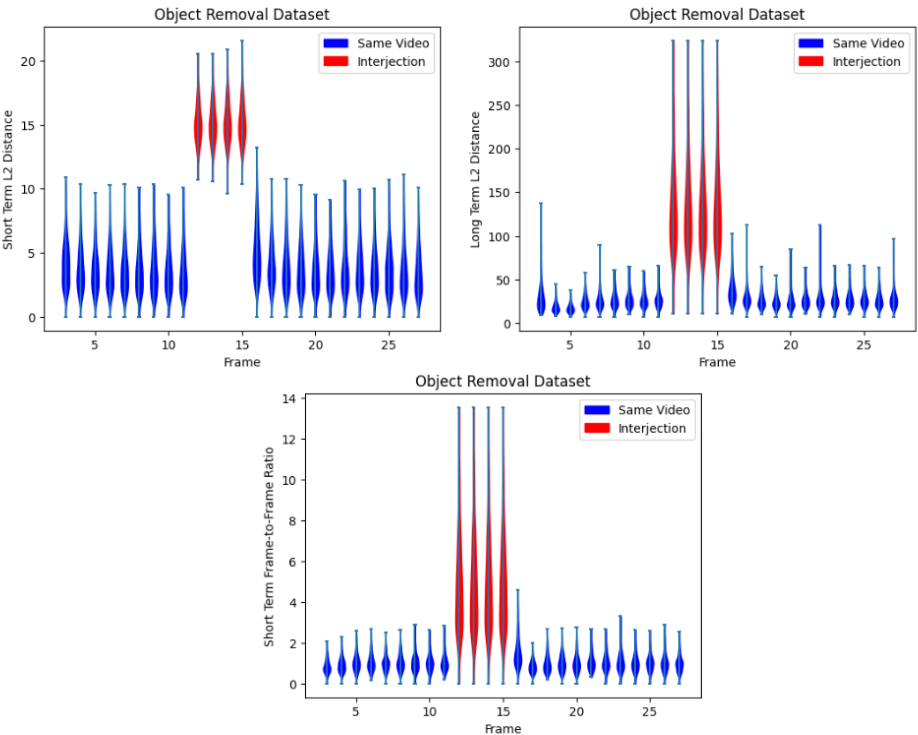}
\caption{\label{fig:pos4_rmo}Object Removal Position 4 Separability}
\end{figure}

Figures \ref{fig:pos4_int} and \ref{fig:pos4_rmo} show that the interjection frames in the object pointer space are extremely separable, even without a linear combination of features. When linearly combining the available features, both the four-frame interjection and object removal datasets should be completely separable.

\subsection{Position 5 - Memory Features}

\begin{figure}[h]
\includegraphics[width=1\columnwidth]{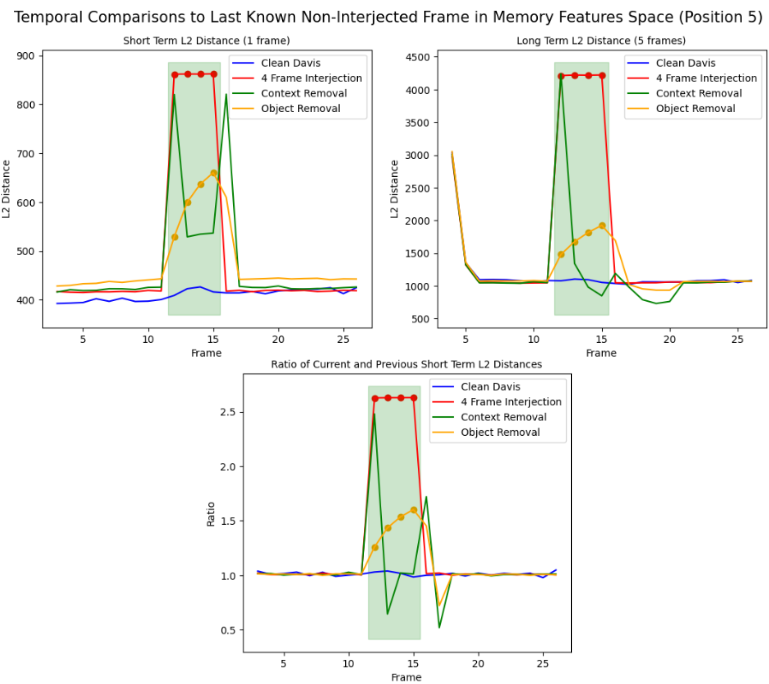}
\caption{\label{fig:pos5_results}Position 5 Features over Time}
\end{figure}

\begin{figure}[h]
\includegraphics[width=1\columnwidth]{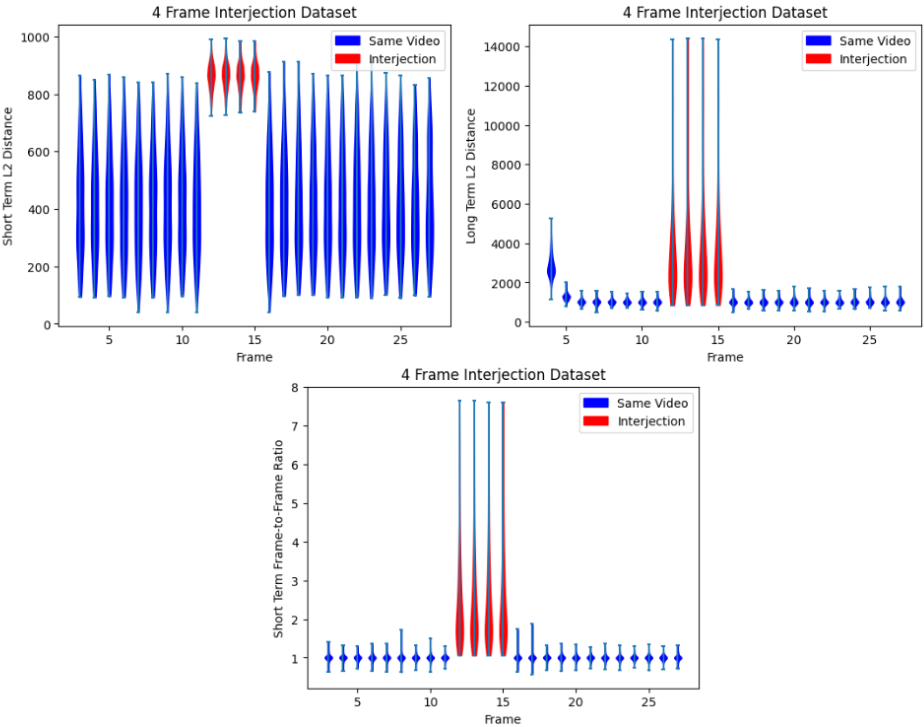}
\caption{\label{fig:pos5_int}Four-Frame Interjection Position 5 Separability}
\end{figure}

\begin{figure}[h]
\includegraphics[width=1\columnwidth]{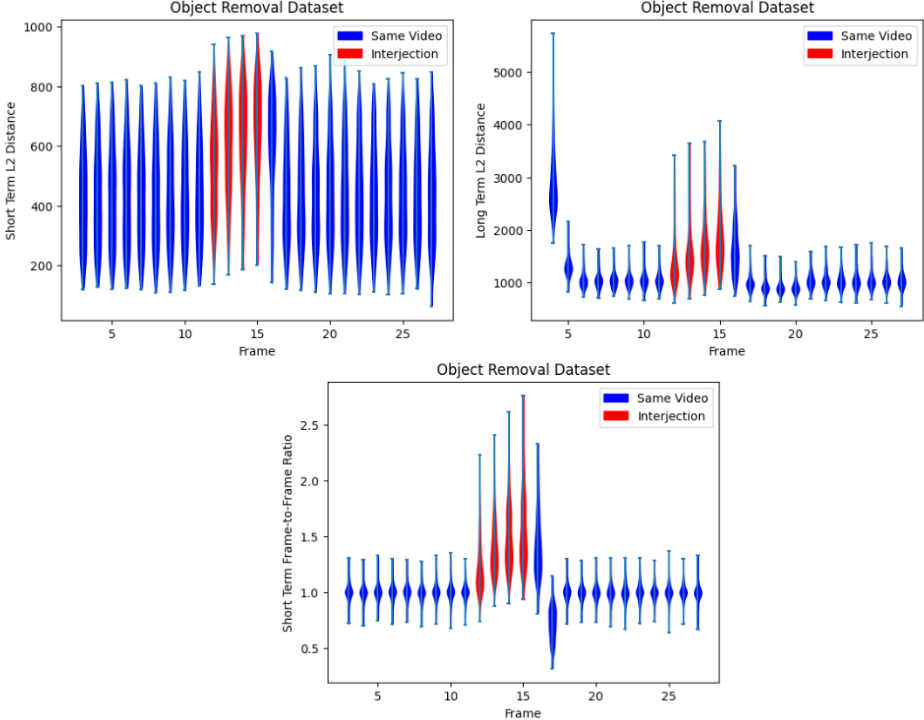}
\caption{\label{fig:pos5_rmo}Object Removal Position 5 Separability}
\end{figure}

Finally, the memory feature space shows results that are significantly worse than those of the object pointer. The dataset plots in Figure \ref{fig:pos5_results} more closely align with positions 1 or 2, demonstrating that the memory contains a representation of the entire image, rather than just the object. The frame distributions in Figures \ref{fig:pos5_int} and \ref{fig:pos5_rmo} are somewhat separable but not nearly as separable at the object pointer.

\section*{Acknowledgment}

This work was performed under the auspices of the U.S. Department of Energy by Lawrence Livermore National Laboratory under Contract DE-AC52-07NA27344 with funding from the lab and the U.S. Navy. Release number: LLNL-JRNL-2002970

\newpage
\bibliographystyle{IEEEtran}

\end{document}